\documentclass[journal]{IEEEtran}

\usepackage{epsfig,amsmath,amssymb,epsf,amsthm,scalefnt,multirow}
\usepackage{xcolor}
\usepackage{float}
\usepackage{cite}
\usepackage{psfrag}
\usepackage{algorithm}
\usepackage{algpseudocode}
\algrenewcommand\algorithmicrequire{\textbf{Input:}}
\algrenewcommand\algorithmicensure{\textbf{Output:}}
\DeclareMathOperator*{\argmax}{argmax}

\newtheorem*{lemma*}{Lemma}

\begin{document}

\title{FCFGS-CV-Based Channel Estimation for Wideband MmWave Massive MIMO Systems with Low-Resolution ADCs}

\author{In-soo Kim and Junil Choi
\thanks{The authors are with the Department of Electrical Engineering, POSTECH, Pohang, Korea (e-mail: insookim@postech.ac.kr; junil@postech.ac.kr).}
\thanks{This work was supported by the Institute for Information \& communications Technology Promotion(IITP) grant funded by the Korea government(MSIT) (No.2016-0-00123, Development of Integer-Forcing MIMO Transceivers for 5G \& Beyond Mobile Communication Systems).}}

\maketitle

\begin{abstract}
In this paper, the fully corrective forward greedy selection-cross validation-based (FCFGS-CV-based) channel estimator is proposed for wideband millimeter wave (mmWave) massive multiple-input multiple-output (MIMO) systems with low-resolution analog-to-digital converters (ADCs). The sparse nature of the mmWave virtual channel in the angular and delay domains is exploited to convert the maximum a posteriori (MAP) channel estimation problem to an optimization problem with a concave objective function and sparsity constraint. The FCFGS algorithm, which is the generalized orthogonal matching pursuit (OMP) algorithm, is used to solve the sparsity-constrained optimization problem. Furthermore, the CV technique is adopted to determine the proper termination condition by detecting overfitting when the sparsity level is unknown.
\end{abstract}

\section{Introduction}
The millimeter wave (mmWave) band in the range of $30$-$300$ GHz offers wide bandwidth, thereby enabling the use of high data rates \cite{6894453}. Furthermore, a large number of antennas can be compactly deployed due to the small wavelength, which is known as massive multiple-input multiple-output (MIMO). One problem of mmWave massive MIMO, however, is the prohibitive analog-to-digital converter (ADC) power consumption incurred by the high sampling frequency and large number of radio frequency (RF) chains. To deal with this issue, one possible solution is to use low-resolution ADCs because the ADC power consumption increases exponentially with the ADC resolution \cite{761034}.

In \cite{8244269, 8320852, 8357527}, compressed sensing-based algorithms for MIMO systems with low-resolution ADCs were proposed. The generalized approximate message passing (GAMP) and vector AMP (VAMP) algorithms \cite{8244269} are low complexity belief propagation-based (BP-based) algorithms. However, GAMP and VAMP break down when the sensing matrix is ill-conditioned. The turbo principle-based generalized expectation consistent signal recovery (GEC-SR) \cite{8320852} and hierarchical model-based sparse Bayesian learning (SBL) \cite{8357527} algorithms are relatively robust to the condition number of the sensing matrix but require matrix inversion, which results in high complexity.

In this paper, we propose the fully corrective forward greedy selection-cross validation-based (FCFGS-CV-based) channel estimator for wideband mmWave massive MIMO systems with low-resolution ADCs. The maximum a posteriori (MAP) channel estimation problem is formulated based on the on-grid compressed sensing framework. To promote the sparsity of the mmWave virtual channel in the angular and delay domains, high grid resolution is employed to reduce the off-grid error between continuous and discretized angles and delays. Then, the problem is converted to an optimization problem with a concave objective function and sparsity constraint. To solve this sparsity-constrained optimization problem, the FCFGS algorithm \cite{doi:10.1137/090759574} is adopted, which is the generalized orthogonal matching pursuit (OMP) algorithm \cite{4385788}. In addition, the CV technique \cite{4301267} is used to detect overfitting, thereby properly terminating FCFGS when the sparsity level is unknown.

\textbf{Notation:} $a$, $\mathbf{a}$, and $\mathbf{A}$ denote a scalar, vector, and matrix. $\|\mathbf{a}\|_{0}$ and $\|\mathbf{a}\|$ represent the $0$-norm and $2$-norm of $\mathbf{a}$. $\|\mathbf{A}\|_{\mathrm{F}}$ is the Frobenius norm of $\mathbf{A}$. The Kronecker product of $\mathbf{A}$ and $\mathbf{B}$ is denoted as $\mathbf{A}\otimes\mathbf{B}$. The support of $\mathbf{a}$ is represented as $\mathrm{supp}(\mathbf{a})$. The row restriction of $\mathbf{A}$ to the index set $\mathcal{I}$ is $\mathbf{A}_{\mathcal{I}}$. $[n]$ denotes $[n]=\{1, \cdots, n\}$. $C([n], k)$ represents the set of all $k$-combinations of $[n]$.

\section{System Model}\label{system_model}
Assume a single-cell uplink massive MIMO system with an $M$-antenna base station and $K$ single-antenna users.\footnote{To compensate the severe path loss in the mmWave band, the users would deploy multiple antennas as well. In practice, however, the limited hardware cost would allow the users to deploy only a single RF chain. Assuming that the users perform proper analog beamforming, the base station would effectively see single-antenna users.} A pair of $B$-bit ADCs is equipped at each receive antenna to reduce the power consumed by the base station. The system operates in the mmWave wideband with $D$ delay taps.

The received signal $\mathbf{y}[n]\in\mathbb{C}^{M}$ at time $n$ is
\begin{equation}
\mathbf{y}[n]=\underbrace{\begin{bmatrix}\mathbf{H}[0]&\cdots&\mathbf{H}[D-1]\end{bmatrix}}_{=\mathbf{H}}\underbrace{\begin{bmatrix}\mathbf{s}[n]\\\vdots\\\mathbf{s}[n-D+1]\end{bmatrix}}_{=\mathbf{s}_{n}}+\mathbf{v}[n]
\end{equation}
where $\mathbf{H}[d]\in\mathbb{C}^{M\times K}$ is the $d$-th delay tap channel, $\mathbf{s}[n]\in\mathbb{C}^{K}$ is the training signal at time $n$ satisfying $\mathbb{E}\{\mathbf{s}[n]\mathbf{s}[n]^{\mathrm{H}}\}=\rho\mathbf{I}_{K}$ with $\rho$ representing the signal-to-noise ratio (SNR), and $\mathbf{v}[n]\sim\mathcal{CN}(\mathbf{0}_{M}, \mathbf{I}_{M})$ is the additive white Gaussian noise (AWGN) at time $n$.

The channels in the mmWave band are characterized by a small number of paths. The $k$-th column of $\mathbf{H}[d]$, which is the channel of the $k$-th user, is \cite{8171203}
\begin{equation}
\mathbf{h}_{k}[d]=\sum_{\ell=1}^{L_{k}}\alpha_{k, \ell}p_{k}(dT-\tau_{k, \ell})\mathbf{a}(\theta_{k, \ell})
\end{equation}
where $L_{k}$ is the number of paths, $\alpha_{k, \ell}\in\mathbb{C}$ is the $\ell$-th path gain, $\theta_{k, \ell}\in[-\pi/2, \pi/2]$ is the $\ell$-th angle-of-arrival (AoA), $\tau_{k, \ell}\in[0, (D-1)T]$ is the $\ell$-th delay, $T$ is the sampling period, $\mathbf{a}(\theta)\in\mathbb{C}^{M}$ is the array response vector, and $p_{k}(t)$ is the pulse shaping filter normalized to satisfy $\mathbb{E}\{\|\mathbf{h}_{k}[d]\|^{2}\}=M$.

In the channel estimation phase of length $N$, the received signal $\mathbf{Y}$ is
\begin{align}
 &\underbrace{\begin{bmatrix}\mathbf{y}[0]&\cdots&\mathbf{y}[N-1]\end{bmatrix}}_{=\mathbf{Y}}\notag\\
=&\mathbf{H}\underbrace{\begin{bmatrix}\mathbf{s}_{0}&\cdots&\mathbf{s}_{N-1}\end{bmatrix}}_{=\mathbf{S}}+\underbrace{\begin{bmatrix}\mathbf{v}[0]&\cdots&\mathbf{v}[N-1]\end{bmatrix}}_{=\mathbf{V}}.
\end{align}
At each receive antenna, the real and imaginary parts of $\mathbf{Y}$ are quantized by a pair of $B$-bit ADCs. The quantized received signal $\hat{\mathbf{Y}}$ is
\begin{equation}\label{quantized_received_signal}
\hat{\mathbf{Y}}=\mathrm{Q}(\mathbf{H}\mathbf{S}+\mathbf{V})
\end{equation}
where $\mathrm{Q}(\cdot)$ is the $B$-bit quantization function defined as
\begin{equation}\label{quantization}
\hat{y}=\mathrm{Q}(y)\iff\begin{cases}\mathrm{Re}(\hat{y}^{\mathrm{lo}})\leq\mathrm{Re}(y)<\mathrm{Re}(\hat{y}^{\mathrm{up}})\\\mathrm{Im}(\hat{y}^{\mathrm{lo}})\leq\mathrm{Im}(y)<\mathrm{Im}(\hat{y}^{\mathrm{up}})\end{cases}
\end{equation}
with $\hat{y}^{\mathrm{lo}}$ and $\hat{y}^{\mathrm{up}}$ the lower and upper thresholds associated with $\hat{y}$. In other words, the real and imaginary parts of $\hat{y}^{\mathrm{lo}}$, $\hat{y}^{\mathrm{up}}$, and $\hat{y}$ correspond to one of the $2^{B}$ quantization intervals.

To promote the sparsity in the angular and delay domains, the virtual channel representation of $\mathbf{H}$ is employed. To transform $\mathbf{h}_{k}[d]$, define the AoA dictionary $\mathbf{B}\in\mathbb{C}^{M\times R_{\mathrm{AoA}}}$ and delay dictionary $\mathbf{p}_{k}[d]\in\mathbb{R}^{R_{\mathrm{delay}}}$ as
\begin{align}
       \mathbf{B}&=\begin{bmatrix}\mathbf{a}(\hat{\theta}_{1})&\cdots&\mathbf{a}(\hat{\theta}_{R_{\mathrm{AoA}}})\end{bmatrix},\\
\mathbf{p}_{k}[d]&=\begin{bmatrix}p_{k}(dT-\hat{\tau}_{1})&\cdots&p_{k}(dT-\hat{\tau}_{R_{\mathrm{delay}}})\end{bmatrix}^{\mathrm{T}}
\end{align}
with $R_{\mathrm{AoA}}\geq M$ discretized AoAs and $R_{\mathrm{delay}}\geq D$ discretized delays. Then, the relationship between $\mathbf{h}_{k}[d]$ and its virtual channel $\mathbf{X}_{k}\in\mathbb{C}^{R_{\mathrm{AoA}}\times R_{\mathrm{delay}}}$ is
\begin{equation}
\mathbf{h}_{k}[d]=\mathbf{B}\mathbf{X}_{k}\mathbf{p}_{k}[d],
\end{equation}
which means that
\begin{equation}
\mathbf{H}[d]=\mathbf{B}\underbrace{\begin{bmatrix}\mathbf{X}_{1}&\cdots&\mathbf{X}_{K}\end{bmatrix}}_{=\mathbf{X}}\underbrace{\begin{bmatrix}\mathbf{p}_{1}[d]&&\\&\ddots&\\&&\mathbf{p}_{K}[d]\end{bmatrix}}_{=\mathbf{P}[d]}
\end{equation}
and
\begin{equation}\label{virtual_channel_representation}
\mathbf{H}=\mathbf{B}\mathbf{X}\underbrace{\begin{bmatrix}\mathbf{P}[0]&\cdots&\mathbf{P}[D-1]\end{bmatrix}}_{=\mathbf{P}}.
\end{equation}
In the sequel, the shorthand notations $L=\sum_{k=1}^{K}L_{k}$ and $R=R_{\mathrm{AoA}}R_{\mathrm{delay}}K$ will be used.

To facilitate the analysis, \eqref{quantized_received_signal} is vectorized in conjunction with \eqref{virtual_channel_representation} as
\begin{equation}\label{vectorized_quantized_received_signal}
\hat{\mathbf{y}}=\mathrm{Q}(\mathbf{A}\mathbf{x}+\mathbf{v})
\end{equation}
where $\hat{\mathbf{y}}=\mathrm{vec}(\hat{\mathbf{Y}})\in\mathbb{C}^{MN}$, $\mathbf{A}=\mathbf{S}^{\mathrm{T}}\mathbf{P}^{\mathrm{T}}\otimes\mathbf{B}\in\mathbb{C}^{MN\times R}$, $\mathbf{x}=\mathrm{vec}(\mathbf{X})\in\mathbb{C}^{R}$, and $\mathbf{v}=\mathrm{vec}(\mathbf{V})\in\mathbb{C}^{MN}$. The goal is to estimate $\mathbf{x}$ from $\hat{\mathbf{y}}$.

\textbf{Remark 1:} The fact which should be emphasized from \eqref{virtual_channel_representation} is that an $L$-sparse $\mathbf{X}$ exists\footnote{In general, $\mathbf{X}$ is not unique as long as $R_{\mathrm{AoA}}>M$ and $R_{\mathrm{delay}}>D$. One possible $\mathbf{X}$ is $\mathbf{X}^{*}=\mathbf{B}^{\mathrm{H}}(\mathbf{B}\mathbf{B}^{\mathrm{H}})^{-1}\mathbf{H}(\mathbf{P}^{\mathrm{T}}\mathbf{P})^{-1}\mathbf{P}^{\mathrm{T}}$, which is the minimum $2$-norm solution.} as $R_{\mathrm{AoA}}\to\infty$ and $R_{\mathrm{delay}}\to\infty$ with $\alpha_{k, \ell}$ as its elements \cite{1033686}. In other words, the off-grid error is negligible when the grid resolution is high.

\section{Proposed Approximate MAP Channel Estimation}
\subsection{FCFGS-Based Channel Estimation}
In this paper, the parameters of $\mathbf{H}$ are assumed to be independent and identically distributed (i.i.d.) as
\begin{align}
\alpha_{k, \ell}&\sim\mathcal{CN}(0, 1),\\
\theta_{k, \ell}&\sim\mathrm{Uniform}([-\pi/2, \pi/2]),\label{aoa_distribution}\\
  \tau_{k, \ell}&\sim\mathrm{Uniform}([0, (D-1)T])\label{delay_distribution}
\end{align}
for all $(k, \ell)$. To build the MAP channel estimation framework, the likelihood function is formulated based on the real counterparts of $\hat{\mathbf{y}}$, $\mathbf{A}$, and $\mathbf{x}$ with the subscript $\mathrm{R}$, which are
\begin{align}
\hat{\mathbf{y}}_{\mathrm{R}}&=\begin{bmatrix}\mathrm{Re}(\hat{\mathbf{y}})^{\mathrm{T}}&\mathrm{Im}(\hat{\mathbf{y}})^{\mathrm{T}}\end{bmatrix}^{\mathrm{T}}\notag\\
                             &=\begin{bmatrix}\hat{y}_{\mathrm{R}, 1}&\cdots&\hat{y}_{\mathrm{R}, 2MN}\end{bmatrix}^{\mathrm{T}},\\
      \mathbf{A}_{\mathrm{R}}&=\begin{bmatrix}\mathrm{Re}(\mathbf{A})&-\mathrm{Im}(\mathbf{A})\\\mathrm{Im}(\mathbf{A})&\mathrm{Re}(\mathbf{A})\end{bmatrix}\notag\\
                             &=\begin{bmatrix}\mathbf{a}_{\mathrm{R}, 1}&\cdots&\mathbf{a}_{\mathrm{R}, 2MN}\end{bmatrix}^{\mathrm{T}},\\
      \mathbf{x}_{\mathrm{R}}&=\begin{bmatrix}\mathrm{Re}(\mathbf{x})^{\mathrm{T}}&\mathrm{Im}(\mathbf{x})^{\mathrm{T}}\end{bmatrix}^{\mathrm{T}}
\end{align}
with the lower and upper thresholds associated with $\hat{\mathbf{y}}_{\mathrm{R}}$ defined as
\begin{align}
\hat{\mathbf{y}}_{\mathrm{R}}^{\mathrm{lo}}&=\begin{bmatrix}\hat{y}_{\mathrm{R}, 1}^{\mathrm{lo}}&\cdots&\hat{y}_{\mathrm{R}, 2MN}^{\mathrm{lo}}\end{bmatrix}^{\mathrm{T}},\\
\hat{\mathbf{y}}_{\mathrm{R}}^{\mathrm{up}}&=\begin{bmatrix}\hat{y}_{\mathrm{R}, 1}^{\mathrm{up}}&\cdots&\hat{y}_{\mathrm{R}, 2MN}^{\mathrm{up}}\end{bmatrix}^{\mathrm{T}}
\end{align}
as in \eqref{quantization}. From now on, the complex and real counterparts are used interchangeably. To proceed with the likelihood function, note that conditioned on $\mathbf{x}$,
\begin{equation}\label{conditional_distribution}
\mathbf{A}\mathbf{x}+\mathbf{v}\sim\mathcal{CN}(\mathbf{A}\mathbf{x}, \mathbf{I}_{MN})
\end{equation}
because $\mathbf{v}\sim\mathcal{CN}(\mathbf{0}_{MN}, \mathbf{I}_{MN})$ is independent of $\mathbf{x}$ as assumed in Section \ref{system_model}. Then, the likelihood function $\ell(\mathbf{x})$ is formulated from \eqref{vectorized_quantized_received_signal} and \eqref{conditional_distribution} as \cite{5456454}
\begin{align}\label{likelihood_function}
\ell(\mathbf{x})&=\mathrm{Pr}\begin{bmatrix}\hat{\mathbf{y}}|\mathbf{x}\end{bmatrix}\notag\\
                &=\prod_{i=1}^{2MN}\left(\Phi\left(\frac{\hat{y}_{\mathrm{R}, i}^{\mathrm{up}}-\mathbf{a}_{\mathrm{R}, i}^{\mathrm{T}}\mathbf{x}_{\mathrm{R}}}{\sqrt{1/2}}\right)-\Phi\left(\frac{\hat{y}_{\mathrm{R}, i}^{\mathrm{lo}}-\mathbf{a}_{\mathrm{R}, i}^{\mathrm{T}}\mathbf{x}_{\mathrm{R}}}{\sqrt{1/2}}\right)\right)
\end{align}
where $\Phi(\cdot)$ denotes the cumulative distribution function (CDF) of $\mathcal{N}(0, 1)$.

To approximate the distribution of $\mathbf{x}$ based on \textbf{Remark 1}, consider the discrete analogue of \eqref{aoa_distribution} and \eqref{delay_distribution}. In other words, consider an $L$-sparse vector with $\mathcal{CN}(\mathbf{0}_{L}, \mathbf{I}_{L})$ as its elements and uniformly distributed discretized AoAs and delays, which is distributed as
\begin{align}
\mathbf{x}_{\mathrm{supp}(\mathbf{x})}&\sim\mathcal{CN}(\mathbf{0}_{L}, \mathbf{I}_{L}),\label{element_distribution}\\
             \mathrm{supp}(\mathbf{x})&\sim\mathrm{Uniform}(C([R], L)).\label{support_distribution}
\end{align}
According to \textbf{Remark 1}, the deviation of \eqref{element_distribution} and \eqref{support_distribution} from the distribution of $\mathbf{x}$ tightens as $R_{\mathrm{AoA}}$ and $R_{\mathrm{delay}}$ increase with the decreasing off-grid error. Therefore, the MAP channel estimation framework is proposed based on \eqref{element_distribution} and \eqref{support_distribution}. The grid resolution is configured as $R_{\mathrm{AoA}}\gg M$ and $R_{\mathrm{delay}}\gg D$ to reduce the off-grid error, which is a widely adopted criterion \cite{7458188}. However, the off-grid error cannot be eliminated unless $R_{\mathrm{AoA}}\to\infty$ and $R_{\mathrm{delay}}\to\infty$. In short, the mismatch between \eqref{element_distribution}, \eqref{support_distribution}, and the distribution of $\mathbf{x}$ can be reduced as the grid resolution is increased but not eliminated.

Now, combining \eqref{likelihood_function}, \eqref{element_distribution}, and \eqref{support_distribution}, the MAP estimator $\hat{\mathbf{x}}$ of $\mathbf{x}$ is formulated as\footnote{The constants independent of $\mathbf{x}$ are neglected for simplicity, which refer to the constants in the Gaussian probability density function (PDF) and probability mass function (PMF) of $\mathrm{Uniform}(C([R], L))$.}
\begin{align}\label{formulation}
\hat{\mathbf{x}}&\overset{\hphantom{(a)}}{=}\argmax_{\mathbf{x}\in\mathbb{C}^{R}}(\ell(\mathbf{x})\cdot e^{-\|\mathbf{x}_{\mathrm{supp}(\mathbf{x})}\|^{2}})\ \text{s.t.}\ \|\mathbf{x}\|_{0}\leq L\notag\\
                &\overset{(a)}{=}\argmax_{\mathbf{x}\in\mathbb{C}^{R}}(\ell(\mathbf{x})\cdot e^{-\|\mathbf{x}\|^{2}})\ \text{s.t.}\ \|\mathbf{x}\|_{0}\leq L
\end{align}
where (a) follows from $\|\mathbf{x}_{\mathrm{supp}(\mathbf{x})}\|=\|\mathbf{x}\|$. From \eqref{formulation}, observe that the objective function is log-concave because $\ell(\mathbf{x})$ has the form of $\Phi(b-x)-\Phi(a-x)$ with $b>a$, which is log-concave, and $e^{-\|\mathbf{x}\|^{2}}$ has the minus of $2$-norm as its exponent, which is concave \cite{5456454}. At this point, define the logarithm of the objective function as $f(\mathbf{x})$ with its gradient $\nabla f(\mathbf{x})$, which are
\begin{align}
       f(\mathbf{x})&=\log\ell(\mathbf{x})-\|\mathbf{x}\|^{2},\\
\nabla f(\mathbf{x})&=\nabla\log\ell(\mathbf{x})-2\mathbf{x}.
\end{align}
Then, the goal is to solve
\begin{equation}\label{map_channel_estimation}
\hat{\mathbf{x}}=\argmax_{\mathbf{x}\in\mathbb{C}^{R}}f(\mathbf{x})\ \text{s.t.}\ \|\mathbf{x}\|_{0}\leq L,
\end{equation}
an optimization problem with a concave objective function and sparsity constraint.

In general, \eqref{map_channel_estimation} is NP-hard to solve because of its sparsity constraint. To approximately solve \eqref{map_channel_estimation} iteratively, the FCFGS algorithm \cite{doi:10.1137/090759574} is adopted, which generalizes the OMP algorithm \cite{4385788} to convex objective functions as presented in Algorithm \ref{fcfgs}. Further explanation of the algorithm is provided in the next subsection but in short, Line 4 selects the largest element of the gradient of the objective function to update the support of the estimate in Line 5. Then, Line 6 performs convex optimization to update the estimate, whose convergence is guaranteed because the objective function and support constraint are convex.

\begin{algorithm}[t]
\caption{FCFGS-CV algorithm to solve \eqref{map_channel_estimation}}\label{fcfgs}
\begin{algorithmic}[1]
\Require $f_{\mathcal{E}}(\cdot)$, $f_{\mathcal{CV}}(\cdot)$
\Ensure $\hat{\mathbf{x}}$
\State $\mathbf{b}=\mathbf{0}_{R}$, $\epsilon=-\infty$
\While {$f_{\mathcal{CV}}(\mathbf{b})>\epsilon$}
\State $\hat{\mathbf{x}}=\mathbf{b}$
\State $i=\displaystyle\argmax_{j\in[R]}|\nabla f_{\mathcal{E}}(\hat{\mathbf{x}})_{j}|$
\State $\mathcal{I}=\mathrm{supp}(\hat{\mathbf{x}})\cup\{i\}$
\State $\mathbf{b}=\displaystyle\argmax_{\mathbf{x}\in\mathbb{C}^{R}}f_{\mathcal{E}}(\mathbf{x})\ \text{s.t.}\ \mathrm{supp}(\mathbf{x})\subseteq\mathcal{I}$
\State $\epsilon=f_{\mathcal{CV}}(\hat{\mathbf{x}})$
\EndWhile
\end{algorithmic}
\end{algorithm}

Before moving on, we mention that the off-grid error incurs the leakage effect \cite{8171203}, which leads to an approximately (instead of exactly) sparse $\mathbf{x}$. As a result, the recovery guarantees of FCFGS established from \cite{doi:10.1137/090759574} break down, which may degrade the accuracy of FCFGS.

\subsection{CV-Based Termination Condition}
In practice, the knowledge of $\|\mathbf{x}\|_{0}$ is critical to determining the proper termination condition of FCFGS. However, $\|\mathbf{x}\|_{0}$ is difficult to acquire because $L$ is often unknown. Furthermore, $\|\mathbf{x}\|_{0}$ may deviate from $L$ when the grid resolution is low. To determine the proper termination condition of FCFGS, the CV technique \cite{4301267} is adopted, which is a model validation technique designed to assess the estimation quality, thereby preventing overfitting.

\begin{figure}[t]
\centering
\includegraphics[width=1\columnwidth]{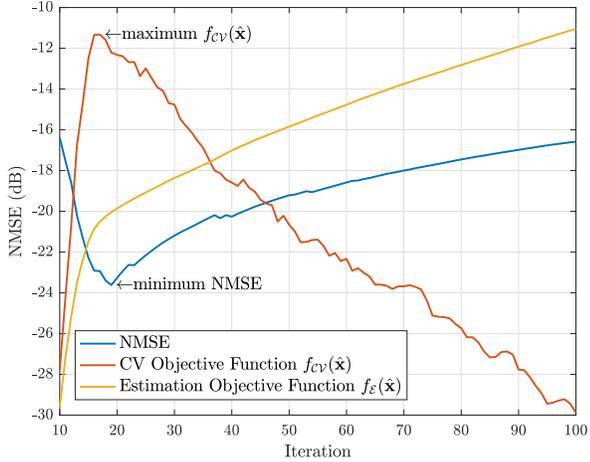}
\caption{NMSE versus iteration of FCFGS-CV with normalized $f_{\mathcal{CV}}(\hat{\mathbf{x}})$ and $f_{\mathcal{E}}(\hat{\mathbf{x}})$ for $B=2$ and $\mathrm{SNR}=0$ dB.}\label{figure_1}
\end{figure}

FCFGS-CV runs estimation using estimation data, whereas validation is performed based on CV data. The disjoint nature of estimation and CV data enables validation to properly assess the quality of the estimation data-based estimate. To proceed, $\hat{\mathbf{y}}$ is divided to the estimation measurement $\hat{\mathbf{y}}_{\mathcal{E}}\in\mathbb{C}^{|\mathcal{E}|}$ and CV measurement $\hat{\mathbf{y}}_{\mathcal{CV}}\in\mathbb{C}^{|\mathcal{CV}|}$ where the disjoint sets $\mathcal{E}$ and $\mathcal{CV}$ partition $[MN]$. The estimation and CV sensing matrices are
\begin{alignat}{2}
&\mathbf{A}_{\mathcal{E}}\in\mathbb{C}^{|\mathcal{E}|\times R}  &&:\ \text{estimation}\ \text{sensing}\ \text{matrix},\\
&\mathbf{A}_{\mathcal{CV}}\in\mathbb{C}^{|\mathcal{CV}|\times R}&&:\ \text{CV}\ \text{sensing}\ \text{matrix}.
\end{alignat}
In addition, define the estimation objective function $f_{\mathcal{E}}(\mathbf{x})$ and CV objective function $f_{\mathcal{CV}}(\mathbf{x})$ using $\hat{\mathbf{y}}_{\mathcal{E}}$, $\mathbf{A}_{\mathcal{E}}$, $\hat{\mathbf{y}}_{\mathcal{CV}}$, and $\mathbf{A}_{\mathcal{CV}}$ as
\begin{align}
 f_{\mathcal{E}}(\mathbf{x})&=\log\mathrm{Pr}\begin{bmatrix}\hat{\mathbf{y}}_{\mathcal{E}}|\mathbf{x}\end{bmatrix}-\|\mathbf{x}\|^{2},\\
f_{\mathcal{CV}}(\mathbf{x})&=\log\mathrm{Pr}\begin{bmatrix}\hat{\mathbf{y}}_{\mathcal{CV}}|\mathbf{x}\end{bmatrix}-\|\mathbf{x}\|^{2},
\end{align}
whose likelihood functions are formulated as in \eqref{likelihood_function}. Then, $\hat{\mathbf{x}}$ is updated based on $f_{\mathcal{E}}(\cdot)$, whereas the quality of $\hat{\mathbf{x}}$ is assessed using $f_{\mathcal{CV}}(\cdot)$. The proposed FCFGS-CV algorithm is presented in Algorithm \ref{fcfgs}. In short, Line 2 tests whether $f_{\mathcal{CV}}(\cdot)$ increases or not to detect overfitting, which is widely used as an indicator of termination. To demonstrate how CV indicates the proper termination timing, define the normalized mean squared error (NMSE) as
\begin{equation}
\mathrm{NMSE}=\mathbb{E}\left\{\frac{\|\hat{\mathbf{H}}-\mathbf{H}\|_{\mathrm{F}}^{2}}{\|\mathbf{H}\|_{\mathrm{F}}^{2}}\right\}
\end{equation}
where $\hat{\mathbf{H}}=\mathbf{B}\hat{\mathbf{X}}\mathbf{P}$, which measures the accuracy of $\hat{\mathbf{x}}$. Then, an illustration of how the NMSE, $f_{\mathcal{CV}}(\hat{\mathbf{x}})$, and $f_{\mathcal{E}}(\hat{\mathbf{x}})$ evolve with the iteration of FCFGS-CV for one problem instance from the simulation results in Section \ref{simulation_results} is shown in Fig. \ref{figure_1}. According to Fig. \ref{figure_1}, the decreasing point of $f_{\mathcal{CV}}(\cdot)$ clearly indicates when the minimum NMSE occurs, whereas $f_{\mathcal{E}}(\cdot)$ provides no clear indication of overfitting.

\begin{figure}[t]
\centering
\includegraphics[width=1\columnwidth]{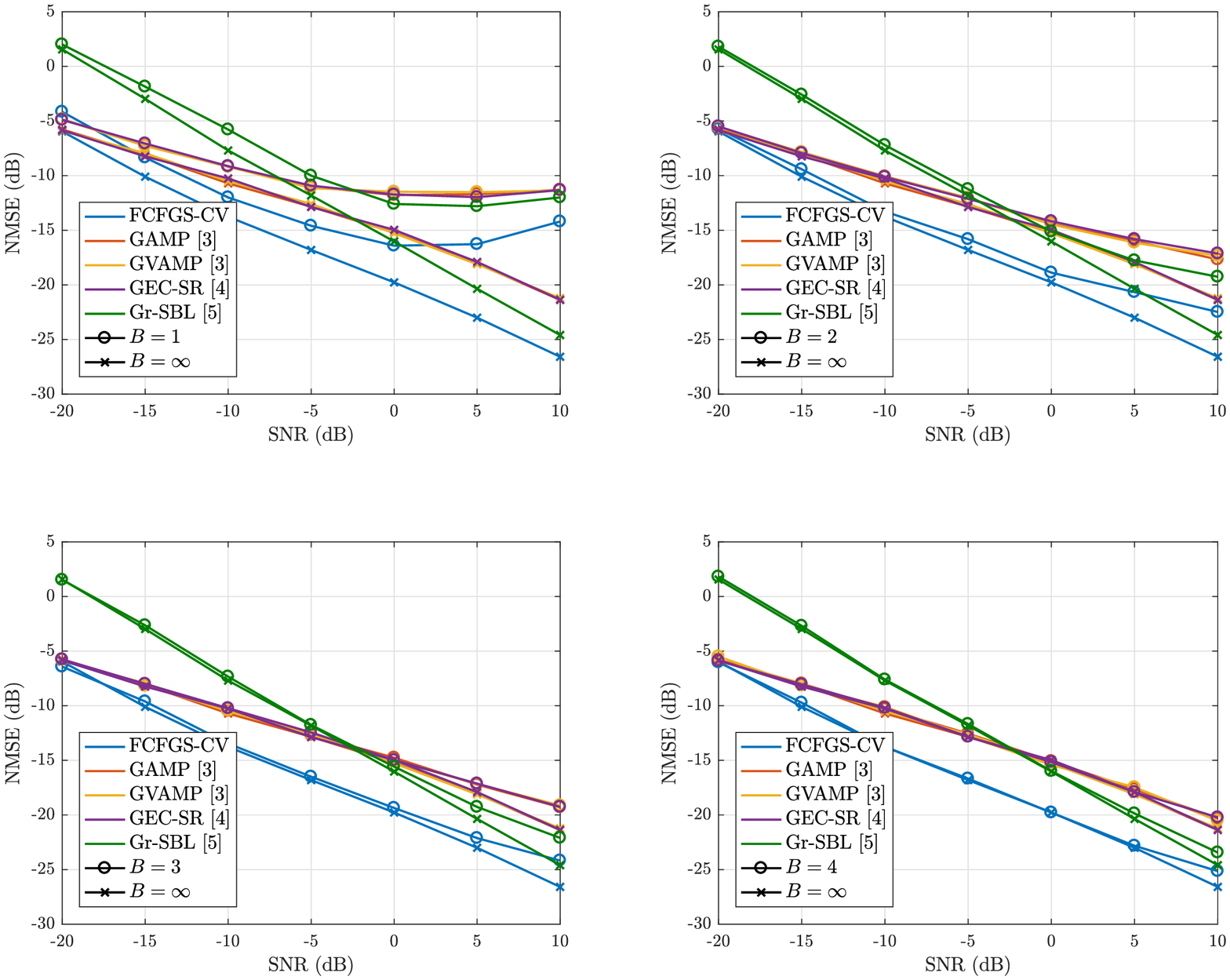}
\caption{NMSE versus SNR for $B=1, 2, 3, 4$ with $N=160$. In each subfigure, $B=\infty$ is shown as a reference.}\label{figure_2}
\end{figure}

\section{Simulation Results}\label{simulation_results}
In this section, we evaluate the performance of the proposed FCFGS-CV-based channel estimator based on its accuracy and complexity. The NMSE is the measure of accuracy. The system parameters are selected as $M=64$ and $K=4$, while $N$ varies from simulation to simulation. The training signals are configured as circularly shifted Zadoff-Chu (ZC) sequences of length $N$ with cyclic prefixes of length $D-1$. The base station employs a uniform linear array (ULA) structure with half-wavelength inter-element spacings, while the users adopt the raised-cosine (RC) pulse shaping filter with a roll-off factor of $0.35$. A pair of $B$-bit uniform quantizers is deployed for the real and imaginary parts at each receive antenna. The channel parameters are $D=8$ with $L_{k}=2$ for all $k$. The dictionaries are configured as $R_{\mathrm{AoA}}=2M$ and $R_{\mathrm{delay}}=2D$ to promote the sparsity of $\mathbf{x}$. To implement CV, among $N$ training signals, $(N-KD)$ signals are used for estimation, while the remaining $KD$ signals are used for CV.

\begin{figure}[t]
\centering
\includegraphics[width=1\columnwidth]{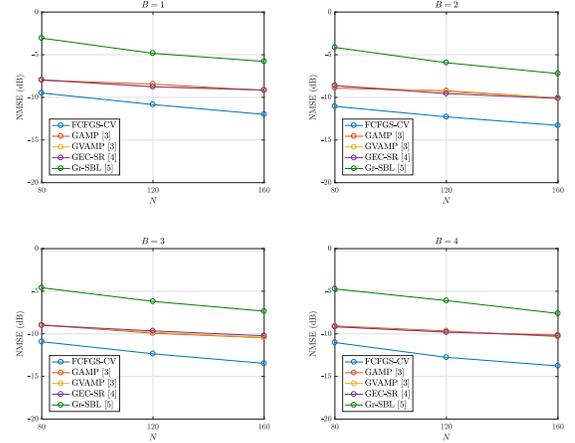}
\caption[caption]{NMSE versus $N$ for $B=1, 2, 3, 4$ with $\mathrm{SNR}=-10$ dB.\\\hspace{0pt}\hphantom{NMSE versus $N$ for $B=1, 2, 3, 4$ with $\mathrm{SNR}=-10$ dB.}}\label{figure_3}
\end{figure}

\begin{figure}[t]
\centering
\includegraphics[width=1\columnwidth]{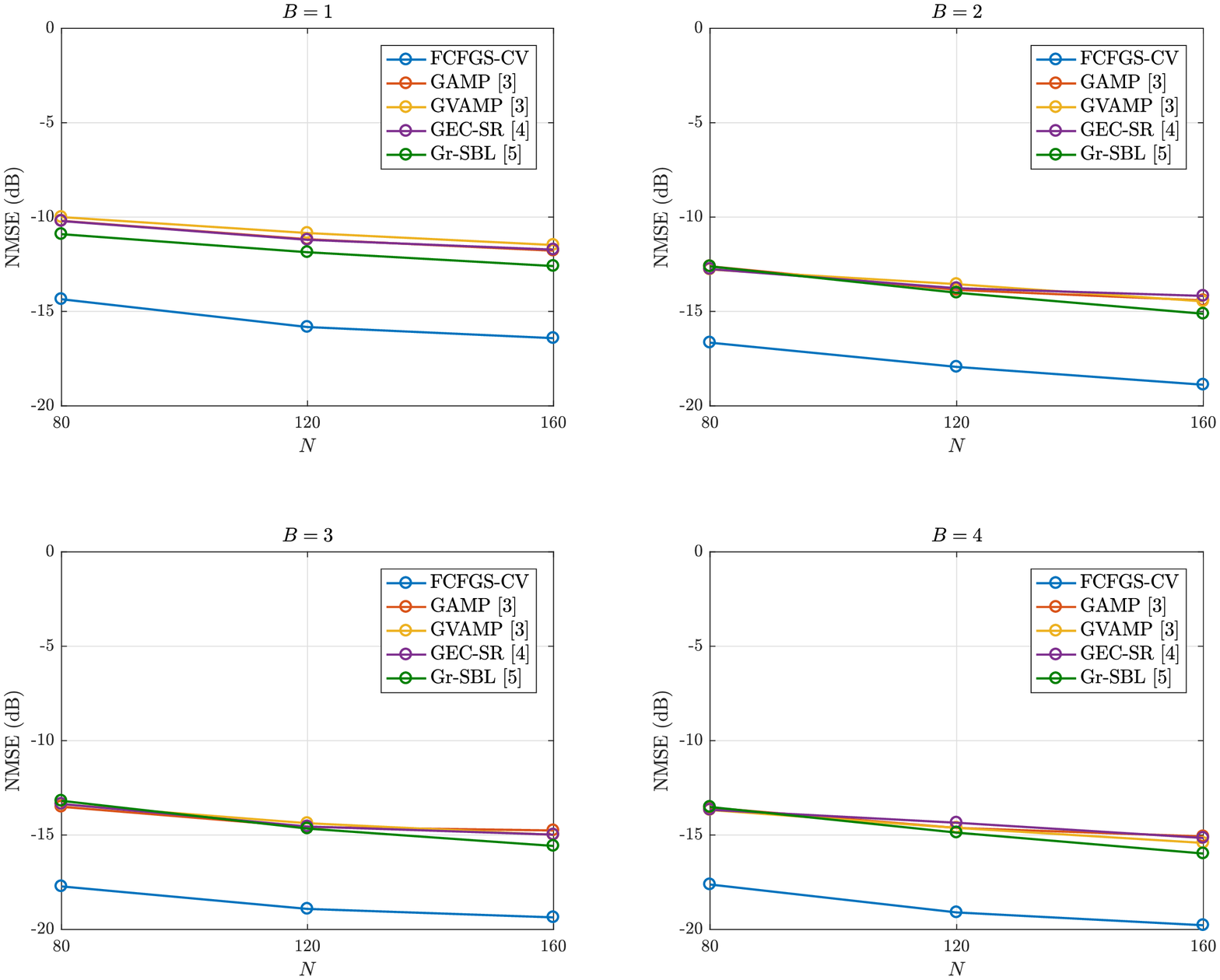}
\caption{NMSE versus $N$ for $B=1, 2, 3, 4$ with $\mathrm{SNR}=0$ dB.}\label{figure_4}
\end{figure}

In the simulation results, the GAMP \cite{8244269}, generalized VAMP (GVAMP) \cite{8244269}, GEC-SR \cite{8320852}, and generalized SBL (Gr-SBL) \cite{8357527} algorithms for generalized linear models (GLMs) \cite{6033942, 7869633} are adopted as benchmarks, which are state-of-the-art compressed sensing-based algorithms. Gr-SBL is configured as $R_{\mathrm{AoA}}=2M$ and $R_{\mathrm{delay}}=2D$. In contrast, GAMP, GVAMP, and GEC-SR are selected as $R_{\mathrm{AoA}}=M$ and $R_{\mathrm{delay}}=D$ because these algorithms diverge when $R_{\mathrm{AoA}}\gg M$ and $R_{\mathrm{delay}}\gg D$ with ill-conditioned sensing matrices.

In Fig. \ref{figure_2}, the NMSEs of FCFGS-CV and other benchmarks are provided for various SNRs with $B=1, 2, 3, 4$ and fixed $N=160$. FCFGS-CV outperforms other benchmarks for all SNRs and $B$ because its high grid resolution guarantees the validity of the assumption established from \eqref{element_distribution} and \eqref{support_distribution}. In contrast, other benchmarks suffer from the mismatch between the postulated distributions (Gaussian mixture and Student-t) and distribution of $\mathbf{x}$. Before moving on, we point out that the NMSE increases in the high SNR regime when $B=1$ for all algorithms because the coarse quantization of $\mathbf{y}$ eliminates the magnitude information of $\mathbf{x}$.

In Figs. \ref{figure_3} and \ref{figure_4}, the NMSEs are analyzed for various $N$ with $B=1, 2, 3, 4$ and fixed $\mathrm{SNR}=-10, 0$ dB. Similar to Fig. \ref{figure_2}, FCFGS-CV shows the best performance among all algorithms for all $N$ and $B$, which reconfirms the superior performance of FCFGS-CV over other benchmarks. Therefore, FCFGS-CV is accurate for not only large $N(=160)$ but moderate $N(=80)$ as well.

Now, the asymptotic complexity of FCFGS-CV is evaluated. Assuming that $R=\mathcal{O}(MDK)$ with $t$-iteration gradient descent method used in Line 6 of Algorithm \ref{fcfgs}, the complexities of the $k$-th iteration of FCFGS-CV, GAMP, GVAMP, GEC-SR, and Gr-SBL are $\mathcal{O}(M^{2}NDK+MNkt)$, $\mathcal{O}(M^{2}NDK)$ \cite{8171203}, $\mathcal{O}(M^{2}NDK)$ \cite{8171203}, $\mathcal{O}(M^{3}ND^{2}K^{2})$ \cite{8320852}, and $\mathcal{O}(M^{3}ND^{2}K^{2})$ \cite{8310593}. The complexities of FCFGS-CV, GAMP, and GVAMP have the same order, whereas the complexities of GEC-SR and Gr-SBL are increased by a factor of $\mathcal{O}(MDK)$ compared to those of FCFGS-CV, GAMP, and GVAMP. Therefore, FCFGS-CV is as efficient as GAMP and GVAMP, which is appealing because AMP-based algorithms are considered as practical channel estimators for massive MIMO systems in the research perspective \cite{8171203}. In practice, the fact that the complexity of FCFGS-CV increases with $M$ as $\mathcal{O}(M^{2})$ may be problematic, so reducing its complexity is left as an interesting future work.

\section{Conclusion}
In this paper, we proposed the FCFGS-CV-based channel estimator for wideband mmWave massive MIMO systems with low-resolution ADCs. The MAP channel estimation framework was formulated as a sparsity-constrained optimization problem with a concave objective function. Then, the FCFGS algorithm was adopted in conjunction with the CV technique, which was revealed to be accurate and efficient in the simulation results.

\bibliographystyle{IEEEtran}
\bibliography{refs_all}

\begin{thebibliography}{10}
\providecommand{\url}[1]{#1}
\csname url@samestyle\endcsname
\providecommand{\newblock}{\relax}
\providecommand{\bibinfo}[2]{#2}
\providecommand{\BIBentrySTDinterwordspacing}{\spaceskip=0pt\relax}
\providecommand{\BIBentryALTinterwordstretchfactor}{4}
\providecommand{\BIBentryALTinterwordspacing}{\spaceskip=\fontdimen2\font plus
\BIBentryALTinterwordstretchfactor\fontdimen3\font minus
  \fontdimen4\font\relax}
\providecommand{\BIBforeignlanguage}[2]{{%
\expandafter\ifx\csname l@#1\endcsname\relax
\typeout{** WARNING: IEEEtran.bst: No hyphenation pattern has been}%
\typeout{** loaded for the language `#1'. Using the pattern for}%
\typeout{** the default language instead.}%
\else
\language=\csname l@#1\endcsname
\fi
#2}}
\providecommand{\BIBdecl}{\relax}
\BIBdecl

\bibitem{6894453}
A.~L. {Swindlehurst}, E.~{Ayanoglu}, P.~{Heydari}, and F.~{Capolino},
  ``Millimeter-wave massive {MIMO}: the next wireless revolution?'' \emph{IEEE
  Commun. Mag.}, vol.~52, no.~9, pp. 56--62, Sep. 2014.

\bibitem{761034}
R.~H. {Walden}, ``Analog-to-digital converter survey and analysis,'' \emph{IEEE
  J. Sel. Areas Commun.}, vol.~17, no.~4, pp. 539--550, Apr. 1999.

\bibitem{8244269}
X.~{Meng}, S.~{Wu}, and J.~{Zhu}, ``A unified bayesian inference framework for
  generalized linear models,'' \emph{IEEE Signal Process. Lett.}, vol.~25,
  no.~3, pp. 398--402, Mar. 2018.

\bibitem{8320852}
H.~{He}, C.~{Wen}, and S.~{Jin}, ``Bayesian optimal data detector for hybrid
  {mmWave} {MIMO}-{OFDM} systems with low-resolution {ADC}s,'' \emph{IEEE J.
  Sel. Topics Signal Process.}, vol.~12, no.~3, pp. 469--483, June 2018.

\bibitem{8357527}
X.~{Meng} and J.~{Zhu}, ``A generalized sparse bayesian learning algorithm for
  1-bit {DOA} estimation,'' \emph{IEEE Commun. Lett.}, vol.~22, no.~7, pp.
  1414--1417, July 2018.

\bibitem{doi:10.1137/090759574}
S.~Shalev-Shwartz, N.~Srebro, and T.~Zhang, ``Trading accuracy for sparsity in
  optimization problems with sparsity constraints,'' \emph{SIAM Journal on
  Optimization}, vol.~20, no.~6, pp. 2807--2832, 2010.

\bibitem{4385788}
J.~A. {Tropp} and A.~C. {Gilbert}, ``Signal recovery from random measurements
  via orthogonal matching pursuit,'' \emph{IEEE Trans. Inf. Theory}, vol.~53,
  no.~12, pp. 4655--4666, Dec. 2007.

\bibitem{4301267}
P.~{Boufounos}, M.~F. {Duarte}, and R.~G. {Baraniuk}, ``Sparse signal
  reconstruction from noisy compressive measurements using cross validation,''
  in \emph{2007 IEEE/SP 14th Workshop on Statistical Signal Processing}, Aug.
  2007, pp. 299--303.

\bibitem{8171203}
J.~{Mo}, P.~{Schniter}, and R.~W. {Heath}, ``Channel estimation in broadband
  millimeter wave {MIMO} systems with few-bit {ADC}s,'' \emph{IEEE Trans.
  Signal Process.}, vol.~66, no.~5, pp. 1141--1154, Mar. 2018.

\bibitem{1033686}
A.~M. {Sayeed}, ``Deconstructing multiantenna fading channels,'' \emph{IEEE
  Trans. Signal Process.}, vol.~50, no.~10, pp. 2563--2579, Oct. 2002.

\bibitem{5456454}
A.~{Mezghani}, F.~{Antreich}, and J.~A. {Nossek}, ``Multiple parameter
  estimation with quantized channel output,'' in \emph{2010 International ITG
  Workshop on Smart Antennas (WSA)}, Feb. 2010, pp. 143--150.

\bibitem{7458188}
J.~{Lee}, G.~{Gil}, and Y.~H. {Lee}, ``Channel estimation via orthogonal
  matching pursuit for hybrid {MIMO} systems in millimeter wave
  communications,'' \emph{IEEE Trans. Commun.}, vol.~64, no.~6, pp. 2370--2386,
  June 2016.

\bibitem{6033942}
S.~{Rangan}, ``Generalized approximate message passing for estimation with
  random linear mixing,'' in \emph{2011 IEEE International Symposium on
  Information Theory Proceedings}, July 2011, pp. 2168--2172.

\bibitem{7869633}
P.~{Schniter}, S.~{Rangan}, and A.~K. {Fletcher}, ``Vector approximate message
  passing for the generalized linear model,'' in \emph{2016 50th Asilomar
  Conference on Signals, Systems and Computers}, Nov. 2016, pp. 1525--1529.

\bibitem{8310593}
Y.~{Ding}, S.~{Chiu}, and B.~D. {Rao}, ``Bayesian channel estimation algorithms
  for massive {MIMO} systems with hybrid analog-digital processing and
  low-resolution {ADC}s,'' \emph{IEEE J. Sel. Topics Signal Process.}, vol.~12,
  no.~3, pp. 499--513, June 2018.

\end{thebibliography}

\end{document}